\documentstyle[epsf,floats,prb,aps]{revtex}
\begin{document}
\draft

\twocolumn[\hsize\textwidth\columnwidth\hsize\csname
@twocolumnfalse\endcsname

\preprint{March 12, 1996}

\title{\bf First-principles study of stability and vibrational
properties of tetragonal PbTiO$_3$}

\author{Alberto Garc\'{\i}a~\cite{byline}}
\address{
Departamento de F\'{\i}sica Aplicada II, Universidad del Pa\'{\i}s
Vasco, Apdo. 644, 48080 Bilbao, Spain}

\author{David Vanderbilt}
\address{Department of Physics and Astronomy, Rutgers University, 
Piscataway, NJ 08855-0849, USA}

\date{March 12, 1996}

\maketitle
\begin{abstract}
A first-principles study of the vibrational modes of PbTiO$_3$ in
the ferroelectric tetragonal phase has been performed at all the
main symmetry points of the Brillouin zone (BZ). The calculations
use the local-density approximation and ultrasoft pseudopotentials
with a plane-wave basis, and reproduce well the available
experimental information on the modes at the $\Gamma$ point,
including the LO-TO splittings. The work was motivated in part by a
previously reported transition to an orthorhombic phase at low
temperatures $\lbrack$J.  Kobayashi, Y.  Uesu, and Y. Sakemi,
Phys. Rev.  B {\bf 28}, 3866 (1983).$\rbrack$ We show that a linear
coupling of orthorhombic strain to one of the modes at $\Gamma$
plays a role in the discussion of the possibility of this phase
transition.  However, no mechanical instabilities (soft modes) are
found, either at $\Gamma$ or at any of the other high-symmetry
points of the BZ.
\end{abstract}
\vskip2pc]

\narrowtext

\section{Introduction}

\begin{figure} 
\epsfxsize=1.7in
\centerline{\epsfbox{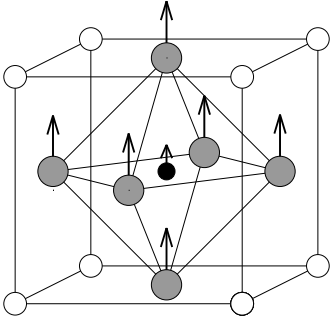}}
\medskip
\caption{Structure of ferroelectric (tetragonal) PbTiO$_3$. The
arrows represent the displacements of the atoms with respect to
their positions in the cubic high-temperature phase. Pb atoms are
depicted by open circles, the Ti atom by the black dot in the
center of the cell, and the O atoms (O$_1$, O$_2$, and O$_3$,
displaced from the Ti atom along $x$, $y$, and $z$, respectively)
by shaded circles.}
\label{fig:cell} 
\end{figure}

Due to their relatively simple structure and the variety of
phenomena they exhibit, the perovskite oxides have become important
subjects of study. Despite sharing a common formula ABO$_3$ and a
highly symmetric high-temperature structure (Fig.~\ref{fig:cell}),
this family of compounds presents a rich and varied low-temperature
phenomenology.  Among the perovskites one finds ferroelectric
crystals such as BaTiO$_3$ and PbTiO$_3$, antiferroelectrics such
as PbZrO$_3$ and NaNbO$_3$, and materials such as SrTiO$_3$ that
exhibit other, non-polar instabilities.

Much progress has been made in the last fifty years in the
experimental characterization of the properties of these
compounds.  One of the main conclusions to emerge from these
studies is the fascinating dependence of the structural and
dynamical behavior on details of chemical composition.  Indeed,
even within a given subgroup of materials one finds significantly
different phase diagrams. For example, BaTiO$_3$ exhibits a
complicated sequence of phase transitions, from cubic to tetragonal
to orthorhombic to rhombohedral, while PbTiO$_3$ shows just one
clearly established transition with $T_c=493^{\circ}$C from the
cubic paraelectric phase to a tetragonal ferroelectric structure.
Moreover, the replacement of Pb for Ba also has important
consequences for the dynamical processes leading to the transition.
It is acknowledged that the soft mode in BaTiO$_3$ is highly
overdamped, and therefore that the transition has some
order-disorder flavor, whereas PbTiO$_3$ has been called a
``textbook example of displacive transition.''~\cite{lines}

Until recently, however, theoretical models of perovskite
properties could not properly take into account the fine chemical
details that distinguish the behavior of the different materials in
this family. Semi-empirical methods are not accurate enough to
model the sort of delicate balance between effects (long-range
dipole interactions vs. short range covalent and repulsion forces,
for example), and schemes based on model Hamiltonians are usually
too simple and too focused on a given material to be of much use in
the unraveling of the chemical trends within the perovskites.

This situation has improved in the last few years with the use of
accurate first-principles density functional calculations to study
the energy surfaces~\cite{cohen,singh,ksv} and even the
temperature-dependent phase diagrams~\cite{zvr-fe,zv-afd,rabe} of
various perovskite oxides. These works have achieved a high degree
of success in reproducing qualitatively and even quantitatively the
experimental observations, giving us confidence that one can now
carry out accurate calculations to elucidate microscopic behavior
(importance of hybridization, competition between long-range and
short-range interactions, etc).  A good example is the recent work
of Rabe and Waghmare,~\cite{rabe} which has helped revise the
conventional wisdom relative to the behavior of PbTiO$_3$.
Indications of a problem with the simple displacive picture were
first seen experimentally in EXAFS measurements,~\cite{exafs} but
the theoretical work\cite{rabe} has provided the microscopic
underpinnings of a partial order-disorder character of the
cubic-tetragonal transition in which the atomic distortions in the
high-temperature phase are proposed to arise from a local
instability.

Another issue, with which we will be mainly concerned in this
paper, is the possible existence of a low-temperature transition.
In the 1950s, Kobayashi {\it et al.\/}~\cite{old-kob} reported the
observation of what appeared to be a distorted (``multiple'')
tetragonal phase of PbTiO$_3$ below approximately $-100{^\circ}$C.
After several negative attempts by other researchers to reproduce
the observations,~\cite{negative} X-ray and optical measurements
were presented~\cite{kob} as corroborating the existence of a
low-temperature phase with an orthorhombic structure. The
transition, at $-90^{\circ}$C, would be second-order, and bring
about a very slight distortion of the tetragonal phase (with the
orthorhombic cell parameters $a$ and $b$ differing by just
4.5$\times 10^{-4}$\AA\ at {$-194^{\circ}$C}) and the direction of
the lattice vectors kept unchanged. The absence of superlattice
reflections would imply a symmetry distortion without
multiplication of the size of the unit cell.

From the point of view of the microscopic dynamics of the
tetragonal
structure, such a transition could be explained by a mechanical
instability of a zone-center phonon whose associated atomic
distortions break the tetragonal symmetry and thus relax the
requirement that $a$ and $b$ be equal. At $T=0$ the energy surface
should then present a saddle point at the configuration
corresponding to the tetragonal phase, with the energy decreasing
along a coordinate representing the amplitude of the soft mode and
the coupled orthorhombic strain.

In this paper we have used first-principles calculations to study
possible mechanical instabilities in the ferroelectric tetragonal
phase of PbTiO$_3$. Our focus has been primarily on homogeneous
(zone-center) distortions of the tetragonal symmetry, aimed at a
detailed theoretical assessment of the possibility of the phase
transition suggested by Kobayashi {\it et al.\/}~\cite{kob}
However, in the interest of completeness, we have also carried out
an analysis of the normal modes at all the main symmetry points on
the surface of the Brillouin zone (BZ). Thus we also present for
the first time a fairly complete collection of normal mode
frequencies and eigenvectors for ferroelectric PbTiO$_3$ computed
from first principles.

The paper is organized as follows. In Sec.~\ref{sec:analysis} we
undertake a classification of the types of possible distortions of
the tetragonal phase of PbTiO$_3$ according to their symmetry.
Sec.~\ref{sec:calcs} briefly describes some technical aspects of
our calculations, whose results are presented in
Sec.~\ref{sec:results}.  Sec.~\ref{sec:conclusions} discusses the
implications of our work for the likelihood of a low-temperature
transition in PbTiO$_3$. The Appendix is devoted to some issues
related to the coupling of atomic displacements to strain degrees
of freedom.

\section{Theoretical analysis of possible
instabilities}\label{sec:analysis}

In the harmonic approximation, the calculation of phonon
frequencies and mode displacement patterns involves the
diagonalization of the dynamical matrix, itself obtained in a
straightforward manner from the force constants
$\Phi^{\alpha\beta}_{ij}$ which enter the expansion of the energy
to second order in the atomic displacements,
\begin{equation}
E = E_0 + \sum_{ij\alpha\beta}
          \Phi^{\alpha\beta}_{ij} u^i_{\alpha}u^j_{\beta} \;\;.
\end{equation}
The force constants can easily be calculated by computing all the
forces caused by a given sublattice displacement.

\begin{table} 
\caption{Character table and decomposition of the vector and
second-order symmetric tensor representations for point group
$4mm$.}
\begin{tabular}{lccccccc}
 &$E$	&$C_4,C_4^{-1}$	&$C_2$	&$m_x,m_y$  &$m_d,m_{d'}$ &V
                                                    &Sym[VxV]\\
\tableline
A$_1$	&1	&1	&1	&1	&1  	&$z$ &
                                               $x^2+y^2,z^2$ \\
A$_2$	&1	&1	&1	&-1	&-1  \\
B$_1$	&1	&-1	&1	&1	&-1     &    &
                                                   $x^2-y^2$ \\
B$_2$	&1	&-1	&1	&-1	&1      &    &$xy$ \\
E	&2	&0	&-2	&0	&0      &$(x,y)$ &
                                                 $(zx,yz)$ \\
\end{tabular}
\label{tab:char}
\end{table}

It is well-known that the normal modes of vibration of a crystal at
a given $k$-point of the BZ transform according to irreducible
representations of the group of the wavevector. Thus a judicious
use of the symmetry information available simplifies the analysis
and saves computational work. Symmetry arguments can also
profitably be used to determine the form of the series expansion of
the total energy of the crystal around a given configuration,
including the correct couplings among various degrees of freedom
(such as atomic displacements and strains). This is precisely what
is needed for a detailed study of the energy surface and the
possible appearance of mechanical instabilities.

In this section we present a brief account of the use of symmetry
considerations to characterize the possible instabilities of the
tetragonal ferroelectric phase of PbTiO$_3$.
Experimentally,~\cite{kob} it has been claimed that the
low-temperature structure has orthorhombic symmetry and there is no
sign of cell doubling. Accordingly, we devote a subsection to the
study of zone-center instabilities of orthorhombic character, and
to the investigation of the form of the energy as a function of the
relevant degrees of freedom.  A second subsection considers
distortions that might conceivably lead to a low-temperature phase
transition but involve a non-orthorhombic symmetry or a doubling of
the unit cell.

\subsection{Orthorhombic instabilities with no cell doubling}

\begin{table}
\caption{Symmetry analysis of the normal modes at different points
of the BZ.}
\begin{tabular}{llll}
{\bf k}, (Group)  &irrep   &no. of copies  &basis  \\
\tableline

$\Gamma$,$Z$ (4mm)\\
     &A$_1$     &4  &Pb$_z$,Ti$_z$,O$_{1z}+$O$_{2z}$,O$_{3z}$ \\
     &B$_1$     &1  &O$_{1z}-$O$_{2z}$  \\
     &E      &5 (2D) &Pb$_x$,Ti$_x$,O$_{1x}$,O$_{2x}$,O$_{3x}$  \\
     &       & 	&Pb$_y$,Ti$_y$,O$_{1y}$,O$_{2y}$,O$_{3y}$   \\
\tableline

$X$,$M'$ (mm2)      \\
     &A$_1$     &5  &Pb$_z$,Ti$_x$,O$_{1z}$,O$_{2z}$,O$_{3x}$   \\
     &A$_2$     &3  &Ti$_y$,O$_{2y}$,O$_{3y}$              \\
     &B$_1$     &2  &Pb$_y$,O$_{1y}$                   \\
     &B$_2$     &5  &Pb$_x$,Ti$_z$,O$_{1x}$,O$_{2z}$,O$_{3z}$    \\
\tableline

$M$,$R$  (4mm)      \\
	&A$_1$	&2	&Pb$_z$,O$_{1y}+$O$_{2x}$	\\
	&A$_2$	&1	&O$_{1x}-$O$_{2y}$	\\
	&B$_1$	&1	&O$_{1y}-$O$_{2x}$	\\
	&B$_2$	&3	&Ti$_z$,O$_{1x}$+O$_{2y}$,O$_{3z}$	\\
	&E	&4 (2D)	&Pb$_x$,Ti$_y$,O$_{2z}$,O$_{3y}$	\\
	&	&	&Pb$_y$,Ti$_x$,O$_{1z}$,O$_{3x}$	\\
\end{tabular}
\label{tab:kptsym}
\end{table}

The ferroelectric phase of PbTiO$_3$ (Fig.~\ref{fig:cell}) is
tetragonal, with space group $P4mm$.  At the $\Gamma$ point, the
group of the wavevector is the point group of the crystal, $4mm$,
characterized by a four-fold rotation axis and four symmetry planes
which contain it.  Table~\ref{tab:char} displays the character
table for $4mm$.  There are five symmetry classes and thus five
irreducible representations (irreps), of which one (E) is
two-dimensional.

The decomposition of the vibrational representation at $\Gamma$ can
be shown by standard techniques to be
\begin{equation}
{\rm Vib}(\Gamma) = 4{\rm A}_1 \oplus {\rm B}_1 \oplus 5{\rm E}
\;\;.
\end{equation}
Physically, this means that the problem of diagonalizing the
$15\times{15}$ dynamical matrix reduces to three simpler tasks: the
diagonalization of a $4\times{4}$ matrix to decouple the four
copies of the A$_1$ irrep, a similar $5\times{5}$ diagonalization
for E, and a simple calculation of a force constant to obtain the
frequency of the B$_1$ mode (its displacement pattern being
completely determined by symmetry). The atomic motions are,
therefore, coupled only within subspaces of the original
fifteen-dimensional configuration space.  The four-dimensional
A$_1$ subspace corresponds to coupled motions with basis [Pb$_z$,
Ti$_z$, O$_{1z}+$O$_{2z}$, O$_{3z}$] and the one-dimensional B$_1$
subspace represents a normal mode with a displacement pattern of
the form [O$_{1z}-$O$_{2z}$]. Of course, at $\Gamma$ there are
three zero-frequency acoustic modes. Two are degenerate (movements
along $x$ or $y$) and transform according to E, and the third is
polarized along $z$ and belongs to A$_1$. The complete symmetry
specification of all the normal modes at $\Gamma$ and at other
high-symmetry $k$-points appears in
Table~\ref{tab:kptsym}.~\cite{aroyo}

It is simple to use this symmetry information to analyze the
possible mechanisms leading to the experimentally suggested phase
transition from the tetragonal to an orthorhombic structure.  By
looking at the $\Gamma$ entry in Table~\ref{tab:kptsym} and
considering the characters in Table~\ref{tab:char}, it is immediate
to conclude that the B$_1$ mode has the right transformation
properties. In this mode the O$_1$ and O$_2$ atoms move in opposite
directions along the $z$ axis, thus breaking the four-fold
symmetry.

\begin{figure}
\epsfxsize=1.8in
\centerline{\epsfbox{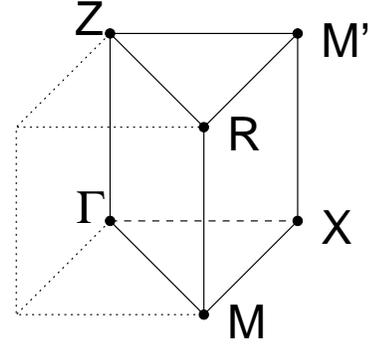}}
\medskip
\caption{Sketch showing the irreducible wedge of the
Brillouin zone associated with the $4mm$ space group, and the
positions of the symmetry points considered in this work.}
\label{fig:bz}
\end{figure}

A calculation of the frequency of this mode is not enough to
determine the existence of an instability, since one should take
into account possible couplings of the atomic displacements to
changes in the size and shape of the unit cell (strain variables).
The possible strains that can be applied to the cell are
represented by the components of a second-order symmetric tensor
($\eta$), and can be classified according to irreducible
representations of the point group of the crystal as shown in the
last column in Table~\ref{tab:char}.  In what follows we use the
notation
\begin{eqnarray}
r&=&\eta_{zz} \;\;, \nonumber\\
s&=&(\eta_{xx}+\eta_{yy})/2 \;\;, \nonumber\\
t&=&(\eta_{xx}-\eta_{yy})/2 \;\;.\nonumber
\end{eqnarray}
Portions of $\eta$ transforming according to the identity
representation A$_1$ leave the tetragonal symmetry unchanged. Such
is the case for $r$ and $s$, which refer to symmetric axial and
in-plane strains, respectively.  The other strain irreps are
associated with lower lattice symmetries: monoclinic for E, and
orthorhombic for B$_1$ and B$_2$. While a B$_2$ ($\eta_{xy}$)
distortion leads to an orthorhombic structure with axes rotated by
45$^{\circ}$ with respect to the tetragonal basis, a pure B$_1$
($t$) strain transforms the cell into an orthorhombic one without a
change in the orientation of the axes.  The latter is precisely the
kind of low-temperature phase suggested for PbTiO$_3$.~\cite{kob}

\begin{table} 
\caption{Character table for the point group $mm2$. The symbols
$m_1$,$m_2$ stand for $m_x$, $m_y$ or $m_d$, $m_{d'}$, depending on
the orientation of the axes.}
\begin{tabular}{lcccc}
 &$E$	&$C_2$	&$m_1$ &$m_y$ \\
\tableline
A$_1$	&1	&1	&1	&1  \\
A$_2$	&1	&1	&-1	&-1 \\
B$_1$	&1	&-1	&1	&-1 \\
B$_2$	&1	&-1	&-1	&1  \\
\end{tabular}
\label{tab:mm2}
\end{table}

Apart from the change in the orientation of the axes, there is an
important difference between B$_2$ ($\eta_{xy}$) and B$_1$ ($t$)
cell distortions.  Since the orthorhombic strain $t$ transforms
according to the B$_1$ irrep, it can couple {\it linearly\/} to the
B$_1$ normal coordinate.~\cite{coupling_ma} Therefore, the crystal
energy expansion considering only the B$_1$ mode and strain is of
the form
\begin{equation}
E = E_0 + {1\over 2} k u^2 + {1\over 2} C t^2 + \gamma u t + ...
  \;\;.
\label{eq:e_b1}
\end{equation}
It is shown in the Appendix that the linear coupling in
Eq.~(\ref{eq:e_b1}) implies a renormalization $C_{\rm eff}= C -
\gamma^2/k$. Thus strain coupling could create instabilities
against B$_1$ (orthorhombic) distortions even if the ``bare''
second order coefficients $k$ and $C$ are positive.

In contrast, any coupling of the B$_2$ strain to a given atomic
displacement $u$ must be at least of second order,
\begin{eqnarray}
E = E_0 + &&{1\over 2} k u^2 + {1\over 2} C \eta_{xy}^2 +
  \nonumber\\
  &&+\gamma u^2 \eta_{xy}^2 + \alpha u^4 + \beta\eta_{xy}^4 + ...
\;\;,
\label{eq:e_b2}
\end{eqnarray}
with no renormalization of the elastic constant $C$ (see
Appendix).

In summary, if the purported low-temperature phase transition in
PbTiO$_3$ is indeed to an orthorhombic phase with no cell doubling,
and with the basis parallel to the tetragonal one, it should be
linked to a negative effective elastic constant $C_{\rm eff}$ for a
$t$ strain. If one allows for the possibility of a rotation of the
axes, the transition could be associated with a negative ``bare''
elastic constant for a B$_2$ strain.

\subsection{Other instabilities}

Apart from the experimentally suggested instability of the
tetragonal phase in favor of an orthorhombic structure with no cell
doubling, there are, in principle, other distortions that might
conceivably lead to phase transitions.

To begin with, and by reference to Table~\ref{tab:char}, one could
think of an instability leading to a phase with monoclinic symmetry
(but still without multiplying the size of the unit cell)
associated with distortions transforming according to the E
irreducible representation. The analysis of this case is
conceptually very similar to the one carried out for the B$_1$
distortions, with the difference that there are eight optical E
modes capable of coupling to strain (four for each of the rows of
the two-dimensional irrep E). Thus $x$- and $y$-polarized normal
modes will couple linearly to $xz$ and $yz$ strains, respectively,
resulting in a renormalized elastic constant $C_{\rm eff}$ for E
distortions.

Next to consider is the possibility of structural phase transitions
associated with a multiplication of the size of the unit cell.
These would come about through the instability of non-$\Gamma$
modes. Since there is no possibility of coupling of these modes to
homogeneous strain at first order, one needs only to compute the
eigenvalues of the force-constant matrix to check for any saddle
points in the energy surface. It is not feasible to study the modes
at all the wavevectors in the BZ, so we focus on a few
high-symmetry $k$-points on the zone surface (see
Fig.~\ref{fig:bz}) which represent cell-doubling distortions.

\begin{table} 
\caption{Structural parameters of PbTiO$_3$.  Theory I and II refer
to a relaxation with constrained lattice constants, and a free
relaxation, respectively.  $z$ atomic coordinates are given in
lattice units. Experimental values are taken from
Ref.~\protect\onlinecite{js_book}.}
\begin{tabular}{lccc}
		&Theory I	&Theory II	&Experiment \\
\tableline

$a$ (a.u.)	&7.380 		&7.298		&7.380 \\
$c/a$		&1.063		&1.054		&1.063 \\
$z$(Ti)		&0.549		&0.537  	&0.540 \\
$z$(O$_1$,O$_2$) &0.630		&0.611		&0.612 \\
$z$(O$_3$) 	&0.125		&0.100  	&0.112 \\
\end{tabular}
\label{tab:struct}
\end{table}

The symmetry analysis of zone-boundary modes proceeds along the
same lines as those for $\Gamma$. Operations that leave the
wavevector invariant will, in general, form subgroups of $4mm$.
For the purposes of our work it suffices to consider just one more
point group, $mm2$, whose character table is given in
Table~\ref{tab:mm2}.~\cite{notation} We show the symmetry
decomposition of atomic displacements at the zone boundary points
in Table~\ref{tab:kptsym}.

\section{Details of calculations}\label{sec:calcs}

The determination of the force constants involves the consideration
of appropriately distorted crystal configurations. Symmetry
arguments are used to reduce to the minimum the number of different
calculations that need to be carried out, and to obtain the
relevant information in the most direct form. For $z$-polarized
modes at the $\Gamma$ point, for example, it is only necessary to
consider the four linearly independent atomic distortions
$(1,0,0,0,0)$, $(0,1,0,0,0)$, $(0,0,1,-1,0)/\sqrt{2}$, and
$(0,0,1,1,-2)/\sqrt{6}$, where the basis is formed by unit $z$
displacements of Pb, Ti, O$_1$, O$_2$, and O$_3$.

Strain parameters are determined by subjecting the crystal to pure
strains and fitting the energy to a polynomial form. The
strain-phonon couplings are computed by finding the forces on the
atoms caused by a suitable strain, since, from Eq.~(\ref{eq:e_b1}),
\begin{equation}
\Bigl({\partial E \over\partial u}\Bigr)_{u=0} = \gamma t \;\;.
\label{eq:gamma}
\end{equation}

We use ultrasoft pseudopotentials, a plane-wave basis set, and a
conjugate-gradients algorithm to compute total energies and forces
for a variety of crystal configurations. The method and the details
of the pseudopotentials employed have been described
elsewhere.~\cite{ksv} For this work we find that a (4,4,4)
Monkhorst-Pack~\cite{mp} sampling of the BZ is enough to provide
good precision in the calculated coefficients (see next section).
Force constants are computed using the Hellmann-Feynman theorem,
with atomic displacements of 0.002 in lattice units.

A final methodological note concerns the calculation of the
frequencies of longitudinal optic (LO) modes at the $\Gamma$
point.  Since our calculations use periodic boundary conditions, we
are not able to introduce a macroscopic electric field, such as it
would arise in an ionic crystal in the presence of a $q\rightarrow
0$ longitudinal vibration. This field creates a splitting of the
frequencies of infrared-active phonons, with the coupling constants
being the ionic effective charges $Z^*$. The force constant matrix
has to be augmented by the effect of a screened (by electronic
effects only) Coulomb interaction among those effective charges,
\begin{equation}
\Phi^{\alpha\beta}_{ij} = \Phi^{\alpha\beta}_{ij}
 + {4\pi e^2\over{\Omega\epsilon_\infty}}Z^*_iZ^*_j
\;\;.
\end{equation}
The effective charges can be obtained from first-principles
calculations.  Here we use those computed for cubic PbTiO$_3$ by
Zhong and Vanderbilt.~\cite{zeffs}

\section{Results}\label{sec:results}

\begin{table} 
\caption{Frequencies of optical modes at $\Gamma$ in cm$^{-1}$.
Infrared-active modes exhibit LO-TO splitting. See text and
Table~\protect\ref{tab:struct} for the meaning of Theory I and
Theory II. Experimental values as compiled in
Ref.~\protect\onlinecite{fk}.}
\begin{tabular}{lddd}
	&Theory I	&Theory II		&Experiment \\
\tableline

A$_1$(TO)	&151	&146	&147  \\
A$_1$(TO)	&355	&337	&359  \\
A$_1$(TO)  	&645	&623	&646  \\
E(TO)		&81 	&82	&88   \\
E(TO)		&183	&195	&220  \\
E(TO)		&268	&237	&289  \\
E(TO)		&464	&501	&505  \\
B$_1$		&285	&280	&289  \\
\tableline

A$_1$(LO)	&187	&186	&189  \\
A$_1$(LO)	&449	&447	&465  \\
A$_1$(LO)	&826	&799	&796  \\
E(LO)		&114	&125	&128  \\
E(LO)		&267	&273	&289  \\
E(LO)		&435 	&418	&436  \\
E(LO)		&625 	&675	&723  \\
\end{tabular}
\label{tab:gamma}
\end{table}

\begin{table}
\caption{Computed frequencies of zone-edge phonons, classified by
symmetry label. The base structure used in the calculations is
Theory I of Table~\protect\ref{tab:struct}. Experimental values are
given when available (Ref.~\protect\onlinecite{fk}).}

\begin{tabular}{llll}
{\bf k}		&irrep	&Frequencies (cm$^{-1}$) &Exp.	\\
\tableline
$Z$        &A$_1$     &102, 189, 447, 831	\\
           &B$_1$     &292	\\
           &E (2)     &46, 151, 184, 270, 454	&59, 168 \\
\tableline
$X$        &A$_1$     &66, 237, 285, 309, 486	&72 \\
           &A$_2$     &131, 233, 426	\\
           &B$_1$     &54, 321	\\
           &B$_2$     &99, 177, 337, 608, 672	\\
\tableline
$M$	&A$_1$	&74, 452	\\
	&A$_2$	&412	\\
	&B$_1$	&138	\\
	&B$_2$	&247, 635, 716	\\
	&E (2)	&57, 203, 294, 398	\\
\tableline
$M'$       &A$_1$     &67, 110, 272, 406, 415	\\
           &A$_2$     &152, 270, 401	\\
           &B$_1$     &57, 329	\\
           &B$_2$     &58, 188, 312, 579, 794	\\
\tableline
$R$	&A$_1$	&90, 411	\\ 	
	&A$_2$	&401	\\
	&B$_1$	&135	\\ 	
	&B$_2$	&200, 626, 803	\\ 	
	&E (2)	&65, 136, 322, 386	\\ 
\end{tabular} 
\label{tab:freq-edge} 
\end{table}

A first concern is the determination of the structural parameters
of the ferroelectric tetragonal phase of PbTiO$_3$.
First-principles LDA calculations typically underestimate the
lattice constants of perovskite oxides by around $1\%$. Our final
objective is the study of dynamical properties of the crystal, and
it would be debatable whether it is better to compute phonon
frequencies and other dynamical parameters at the experimental or
at the theoretical lattice constant.  Past experience with
perovskites has shown that the displacement patterns associated
with some soft modes, and even the existence of the latter, depend
on lattice constant and strain.~\cite{cohen,singh} In the case of
ferroelectric PbTiO$_3$ there is an additional complication, namely
the existence of internal atomic displacements, which are of course
coupled to the cell dimensions. Our first strategy was to use the
experimental lattice constants $a$=7.380 a.u., $c/a=1.0635$ and
optimize the internal atomic positions to obtain a base reference
configuration with zero forces with which to compute phonon
frequencies and strain coefficients.  We call this ``Theory I.''
Later we determined an optimized structure (cell shape and atomic
positions coupled) via a special minimization procedure (see
Appendix); we call this ``Theory II.''  Table~\ref{tab:struct}
summarizes the structural information.  While, as we shall see, we
obtain substantially the same phonon frequencies in either case,
the second approach, using as a reference the structure which gives
a theoretical energy minimum with respect to strain, is in
principle more appropriate for the calculation of elastic
properties and strain-phonon couplings.

As part of the investigation of the possible mechanical
instabilities,~\cite{dynamical} we have obtained a complete set of
calculated phonon frequencies for PbTiO$_3$. These are given, along
with experimental results when available, in Tables~\ref{tab:gamma}
and \ref{tab:freq-edge}.~\cite{more} The agreement of our
theoretical results with experiment for the zone-center modes (both
TO and LO) is quite good. We are thus confident that our
computational approach can be trusted in its predictions of
zone-edge vibrational frequencies that have not yet been determined
experimentally.  To our knowledge, the only other calculation of
vibrational frequencies and modes for tetragonal PbTiO$_3$ was
carried out by Freire and Katiyar.~\cite{fk} An important
difference with our work is that those authors used an empirical
fitting procedure to adjust the parameters of a rigid-ion model. We
use no empirical parameters of any kind, just the atomic numbers
and masses of the atoms involved.  Table~\ref{tab:gamma} can be
used also to estimate the degree of dependence of the phonon
frequencies upon the details of the base structure used in the
calculations (``Theory I'' or ``Theory II'' above).  Phonons at
zone-boundary points are computed using the ``Theory I''
structure.

\begin{table} 
\caption{Test of the convergence of mode frequencies with
$k$-point grid. (4,4,4) and (6,6,6) grids are used for
the ``Theory I'' choice of Table~\protect\ref{tab:struct}. 
The frequencies (in cm$^{-1}$) are those of the transverse
$z$-polarized modes at $\Gamma$.}

\begin{tabular}{lcc}
$k$-point grid	&(4,4,4)     	&(6,6,6)	\\
\tableline
A$_1$(TO)		&151		&153	\\
B$_1$			&285		&289	\\
A$_1$(TO)		&355		&359	\\
A$_1$(TO)		&645		&648	\\
\end{tabular}
\label{tab:test}
\end{table}

To test the convergence of our results with respect to the density
of the $k$-point grid for BZ integrations, we recomputed the
frequencies of $z$-polarized $\Gamma$ modes using a (6,6,6)
Monkhorst-Pack grid. The results, displayed in
Table~\ref{tab:test}, indicate a high level of convergence.

As for the question of the existence of a phase transition at low
temperature, we find that all the vibrational frequencies are real,
as can be seen from the positive sign of all the mode force
constants $k$. Thus there are no mechanical instabilities in the
``bare'' vibrational degrees of freedom, either at $\Gamma$ or at
the edges of the BZ. However, there still remains the question of
whether the linear coupling to strain degrees of freedom could
result in any instability.

\begin{figure} 
\epsfxsize=2.8in
\centerline{\epsfbox{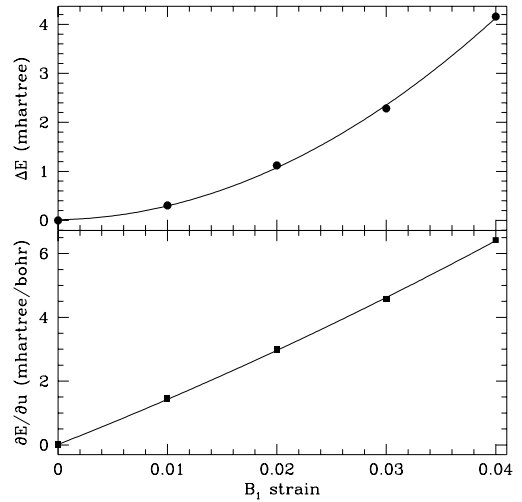}}
\medskip
\caption{Upper panel: Change in crystal energy (per cell) as a
function of the orthorhombicity parameter $t$ (B$_1$ strain). The
curve is a fit to a parabola. Lower panel: Derivative of the
crystal energy with respect to the B$_1$ normal mode amplitude at
zero amplitude, as a function of B$_1$ strain. According to
Eq.~(\protect\ref{eq:gamma}), it measures the degree of coupling
between the normal mode and the strain.}
\label{fig:b1_fit} 
\end{figure}

We deal first with the renormalization of the elastic constant
corresponding to a B$_1$ orthorhombic strain.  By applying pure
B$_1$ strains of different magnitudes (for which we set $b-a\neq 0$
while keeping the sum $a+b$ constant) and computing the resulting
values of the total energy, we obtain the data plotted in
Fig.~\ref{fig:b1_fit}. A fit to a simple parabola is very good up
to sizable strains.  The elastic constant $C$ [see
Eq.~(\ref{eq:e_b1})] turns out to be 5.0 hartree.~\cite{units} As
mentioned above, we use the optimized structure (``Theory II'') for
this and the rest of the calculations involving elastic constants
and strain-phonon couplings.

From the same set of calculations, but extracting this time the
forces
on the atoms and taking the scalar product (in configuration space)
with the eigenvector of the B$_1$ mode, we obtain from
Eq.~(\ref{eq:gamma}) (see also Fig.~\ref{fig:b1_fit}) $\gamma=0.15$
hartree/bohr. The force constant for the B$_1$ mode is 0.048
hartree/bohr$^2$, so the renormalized $C$ is $C_{\rm eff}$= 4.5
hartree. We see that even though there is a 10\% change in the
value of the elastic constant, the renormalization due to the
coupling to the phonons is not enough to cause a B$_1$ instability
of the tetragonal cell.

We performed a similar set of calculations for the analysis of the
monoclinic distortion with E symmetry.  The forces along the $x$
axis appearing upon application of an $\eta_{xz}$ strain translated
into coupling constants of 0.17, 0.05, 0.06, and 0.00 hartree for
the optical $x$-polarized E modes of respective force constants
0.014, 0.042, 0.077, and 0.155 hartree/bohr$^2$.  The bare elastic
constant for $\eta_{xz}$ strain is 5.4 hartree. Adding up the
contributions to the renormalization from the four modes we
obtained an effective elastic constant $C_{\rm eff}$ of 3.3
hartree.  In this case the renormalization amounts to 40\% of the
bare value, but still is not enough to drive an E instability.

As discussed above, there is no linear coupling of B$_2$
orthorhombic strains to atomic displacements. The calculated
elastic constant for this type of strain is positive (6.0 hartree),
so there should be no instabilities of B$_2$ symmetry either.

Finally, recall that there is no first-order coupling of
zone-boundary modes to homogeneous strain.  Thus we need only check
the bare force constants, which are all found to be positive (see
Table~\ref{tab:freq-edge}). This means that we do not expect any
mechanical instabilities associated with a cell doubling.

\section{Conclusions}\label{sec:conclusions}

The low-temperature transition proposed by Ko\-ba\-ya\-shi {\it et al.}
\cite{kob} on the basis of X-ray and optical measurements is
supposed to involve a slight orthorhombic distortion of the
tetragonal phase, maintaining the orientation of the cell axes with
no cell doubling.  Our analysis of the energetics of B$_1$
distortions shows that a low-temperature transition of this kind is
possible in principle, but not likely in ferroelectric PbTiO$_3$.
In this connection, it should be noted that, to our knowledge, the
experimental observations of Ref.~\onlinecite{kob} have not been
reproduced since 1983.

We also checked more generally for other kinds of low-temperature
structural transitions.  However, we find that all
unit-cell-preserving distortions exhibit positive elastic
constants, thus apparently ruling out transitions to a monoclinic
structure (E distortions) or to a 45$^\circ$-rotated orthorhombic
structure (B$_2$ strain). Furthermore, we show that there are no
mechanical instabilities associated with zone-boundary normal modes
that could cause a phase transition with cell doubling.

Since we have not exhaustively explored the vibrational spectrum of
the crystal, it is conceivable that a mechanical instability at a
$k$-point not on the BZ boundary may have been missed.  However,
our work shows fairly clearly that a simple transition is not
likely in ferroelectric PbTiO$_3$ at low temperatures.

\acknowledgments

This work was supported in part by the ONR Grant N00014-91-J-1184
and by the UPV research grant 060.310-EA149/95.  Thanks are due to
J.M.  P\'erez-Mato, M. Aroyo, W.  Zhong, and U. Waghmare for useful
comments.

\appendix
\section*{}

\subsection{Renormalization of energy-surface coefficients}

We show first how the linear coupling of $u$ to $t$ in the energy
expansion of Eq.~(\ref{eq:e_b1}) implies a renormalization of $C$
(or, equivalently, of $k$). After a transformation of the quadratic
form to ``principal axes'' by a linear change of variables, the
first partial derivatives of the energy will be zero. We can
achieve the transformation implicitly by setting the derivative of
$E$ with respect to $u$ to zero, and solving for $u$, to get
$u=-\gamma t/k$.  When this condition is inserted back into
Eq.~(\ref{eq:e_b1}), we obtain an expression for E as a function of
the free variable $t$,
\begin{equation}
E(t) = {1\over{2}}(C-{\gamma^2\over{k}})t^2 
= {1\over{2}}C_{\rm eff}~t^2 \;\;,
\end{equation}
from which it follows that the effective elastic constant is
$C_{\rm eff}= C - \gamma^2/k$. (If instead $u$ is chosen as a free
variable,  one obtains a renormalized spring constant $k_{\rm eff}
= k - \gamma^2/C$.  However, the physical mode frequency is not
renormalized, because of the ``infinite mass'' associated with the
strain degrees of freedom.)

In the case of the B$_2$ distortion with quadratic coupling,
Eq.~(\ref{eq:e_b2}), one needs $\beta>0$ and $\alpha >0$ or else
there would be unphysical divergences to $-\infty$ in the energy.
But then, setting the $u$-derivative of the energy to zero, one
gets either $u=0$ (trivial) or $u^2=-(k+2\gamma t^2)/{4\alpha}$
(meaningless since $u$ would be imaginary).  Thus there is no
renormalization of the elastic constant.

\subsection{Optimization of structural parameters}

Using the symmetry constraints of the $4mm$ point group, one can
write down the expression (to second order in the strain and atomic
displacements) for the energy of a general tetragonal phase of that
symmetry as
\begin{equation}
E=E_0 + E_{\rm strain} + E_{\rm internal} + E_{\rm strain-ph} \;\;,
\end{equation}
where 
\begin{equation}
E_{\rm strain}=
   \alpha_1 s +\beta_1 r + \alpha_2 s^2 + \beta_2 r^2 + \delta sr
\end{equation}
is the part that depends only on the $s$ and $r$ strains,
\begin{equation}
E_{\rm internal} = \sum_{i=1}^{3}{{1\over 2}k_i u_i^2}
\end{equation}
is the change in energy due to internal atomic displacements
compatible with the symmetry (and thus expanded as combinations of
the three A$_1$ phonons polarized along the $z$ axis), and
\begin{equation}
E_{\rm strain-ph} = \sum_{i=1}^{3}{(\gamma_s^i u_i s +
  \gamma_r^i u_i r)}
\end{equation}
are the symmetry-allowed couplings of $s$ and $r$ to the A$_1$
phonons (both $s$ and $r$ transform according to A$_1$).

The fourteen coefficients in this expansion are easily computed for
a given base configuration. In our case, the starting point is a
tetragonal cell with $a$ and $c$ given by experiment and the
internal atomic positions along the $z$ axis optimized
theoretically to eliminate residual forces (column labeled ``Theory
I'' in Table~\ref{tab:struct}). Computed A$_1$ phonon frequencies
directly give the force constants $k_i$, and the strain and
strain-coupling coefficients are obtained in a manner analogous to
that described in the main body of the paper. Once the quadratic
form for $E$ is known, it is a simple matter to find the structural
parameters which correspond to the minimum energy (column labeled
``Theory II'' in Table~\ref{tab:struct}). As is typical of
first-principles calculations, the calculated lattice parameters
are smaller than the experimental values by around $1\%$.


\begin{references}
\bibitem{notation} Since the notation for irreducible
representations at different points of the BZ is not standardized,
we prefer to use the customary irrep labeling for point groups (A,
B, E, etc.) and refer to specific representations by mentioning the
$k$-point explicitly.
\bibitem{dynamical} For the purpose of discussing mechanical
instabilities, one does not need to take into account the masses of
the atoms; it suffices to analyze the eigenvectors and eigenvalues
of the force-constant matrix. However, once this matrix is
determined, it is straightforward to include the relevant mass
factors to obtain the dynamical matrix and phonon frequencies (and
to augment the dynamical matrix with the long-range coulomb
coupling terms to get the LO-TO splitting).
\bibitem{more} The mode eigenvectors are not listed, but can be
obtained from the authors upon request.
\bibitem{units} Our energies in Eqs.~\ref{eq:e_b1} and
\ref{eq:e_b2} are referred to a fixed cell volume. It is customary
to give elastic constants in units of energy/volume; we are
implicitly using units of energy/cell.
\end{references}
\end{document}